\documentclass[twocolumn,prl,superscriptaddress,longbibliography]{revtex4-1}
\usepackage{multirow}
\usepackage{graphicx}
\usepackage{amsmath}
\usepackage{enumerate}
\usepackage{epstopdf}
\usepackage{times}
\usepackage{color}
\usepackage{bm}

\usepackage[colorlinks,bookmarks=false,citecolor=blue,linkcolor=red,urlcolor=blue]{hyperref}

\begin{document}

\title{Destroying a topological quantum bit by condensing Ising vortices}
\author{Zhihao Hao}
\email{zhhhaolong@gmail.com}
\affiliation{Department of Physics and Astronomy, University of Waterloo, Waterloo, ON, N2L 3G1, Canada}
\author{Stephen Inglis}
\affiliation{Department of Physics and Astronomy, University of Waterloo, Waterloo, ON, N2L 3G1, Canada}
\affiliation{Department of Physics, Arnold Sommerfeld Center for Theoretical Physics and Center for NanoScience, University of Munich, Theresienstrasse 37, 80333 Munich, Germany}
\author{Roger Melko}
\affiliation{Department of Physics and Astronomy, University of Waterloo, Waterloo, ON, N2L 3G1, Canada}
\affiliation{Perimeter Institute for Theoretical Physics, Waterloo, Ontario N2L 2Y5, Canada}
\begin{abstract}
The imminent realization of topologically-protected qubits in fabricated systems
will provide not only an elementary implementation of fault-tolerant quantum computing 
architecture, but also an experimental vehicle for the general study of 
topological order. 
The simplest topological qubit harbors what is known as a Z$_2$ liquid phase, 
which encodes information via a degeneracy depending on the system's topology. 
Elementary excitations of the phase are
fractionally charged objects called  spinons, or Ising flux vortices called visons.
At zero temperature a Z$_2$ liquid is stable under deformations of the Hamiltonian 
until spinon or vison condensation induces a quantum phase transition destroying the topological order.
Here, we use quantum Monte Carlo to study a vison-induced transition from a Z$_2$ liquid to a valence-bond solid in a quantum dimer model on the kagome lattice.  Our results indicate that this critical point is beyond the description of the standard Landau paradigm.
\end{abstract}
\maketitle

\section{Introduction}
A quantum bit (qubit) is the basic unit for quantum information processing and quantum computation.  Among different proposed realizations, a topologically-protected qubit is one of the most attractive candidates thanks to its inherent stability against decoherence. 
Unlike conventional qubits, no material or physical system is known to harbor a state appropriate to construct a topological
qubit.  However, several groups have recently outlined a strategy \cite{DiVincenzo,Fowler1} for artificially fabricating one elementary example, called a Z$_2$ topological
qubit, culminating in the recent proof-of-principle implementation with three \cite{transmon} and five \cite{transmon2} coupled superconducting transmon qubits.

Although actual quantum computation can only be possible if thousands of qubits are amassed and properly assembled, the potential creation of a single Z$_2$ topological qubit would open the door to study the properties of a topologically ordered system experimentally, for example its feasibility as quantum memory \cite{TopoMem}.
In light of this, and to serve as motivation for the perfection of fabrication and characterization techniques, it is 
highly desirable to identify interesting physics that exist at the level of one qubit.
An important class of problems concerns how a logical qubit, protected by Z$_2$ topological order, is destroyed by two types of errors in the physical qubits
used to construct it, conventionally referred to as ``bit-flips'' and ``phase-flips''.

Physically, a logical qubit protected by Z$_2$ topological order can emerge from the low-energy subspace of a many-body Hamiltonian \cite{ToricCode}. The qubit encodes information through the $2^{g}$ degeneracy of its ground state where $g$ is the genus of the hosting manifold. 
The simplest realization is a quantum spin liquid characterized by a Z$_2$ gauge theory with electric field and magnetic fluxes defined up to modulus two \cite{LeonQSL}.  
The bit-flip and phase-flip errors of the physical qubits correspond to two
low energy excitations of the liquid: spinons carrying electric charges and visons possessing magnetic fluxes, respectively. 
Remarkably, it is predicted that these excitations, or errors, may condense at $T=0$,  leading to 
novel continuous quantum phase transitions described by critical theories beyond the paradigm of conventional Landau theory \cite{DQCP,Xu_1}.
The continuous phase transition caused by condensing spinons has been previously studied \cite{Isakov2012}. In contrast, while a study \cite{Ralko2} of vison condensation transition by directly probing vison dynamics exists,  the critical properties of the transition involving visons have not been investigated in a quantum many-body model to date. 

Recently, pioneering work by several groups \cite{Yan.2011,Depenbrock.2012,Jiang2012,Jiang2012_2} established that a Z$_2$ spin liquid is the ground state of a set of very simple models in frustrated magnetism. The liquid can be viewed as a coherent fluid of short-range spin singlets, commonly referred to as ``dimers'' or ``valence bonds''. The condensation of visons leads to phases with spatially modulated dimer densities, or valence-bond solids (VBS). In particular, it has been demonstrated \cite{Yan.2011,Depenbrock.2012,Jiang2012} that the spin-1/2 antiferromagnet on the kagome lattice hosts a ground state with Z$_2$ topological order with several VBS \cite{singh.2008,Nikolic.2003} adjacent. It thus constitutes a good candidate to study the universal properties promoted by a phase transition where the topological phase is destroyed by condensing visons.

While the Density Matrix Renormalzation Group (DMRG) and related tools \cite{Yan.2011,Depenbrock.2012,Jiang2012} are sufficent to identify some features of the Z$_2$ topological liquid in quasi-2D kagome lattices \cite{ARCMP_Miles}, fully characterizing quantum phase transitions out of the liquid phase is not feasible using these methods due to the large lattice sizes required as the correlation length diverges.  The ideal method for the task is scalable simulation
methods such as quantum Monte Carlo (QMC) \cite{ARCMP_Kaul}.  Since a direct QMC simulation of the spin model \cite{Yan.2011} is impossible due to the existence of the infamous sign problem \cite{ARCMP_Kaul}, we simulate a quantum dimer model (QDM) on the kagome lattice, which contains a stable Z$_2$ liquid phase over a large region of its phase diagram.  After mapping out the phase diagram via large-scale $T=0$ diffusive QMC \cite{Buonaura.1998}, we examine the quantum phase transitions out of the liquid to two VBS phases. 
For the transition between one of the VBS phases and the Z$_2$ liquid, 
our QMC simulations provide the first numerical evidence for novel quantum critical properties
driven by vison condensation.  
These critical properties are consistent with the exotic O(4$^\ast$) universality class \cite{Xu_1} previously studied theoretically \cite{Calabrese.2003} and numerically \cite{Isakov.2005} in
effective field theories.
\begin{figure}
\centering
\includegraphics[width=\columnwidth]{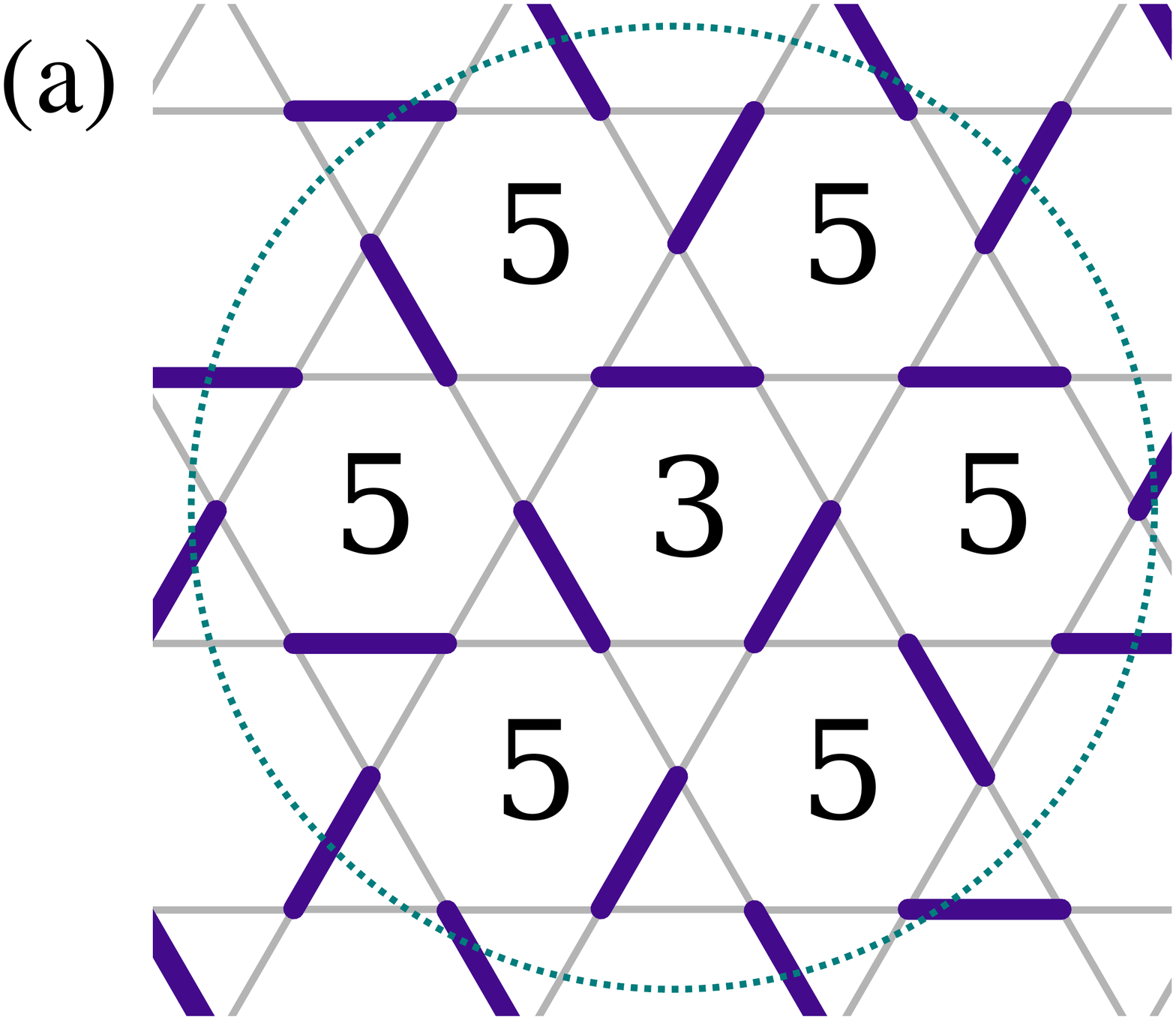}
\newline \vspace{0.5cm}
\includegraphics[width=\columnwidth]{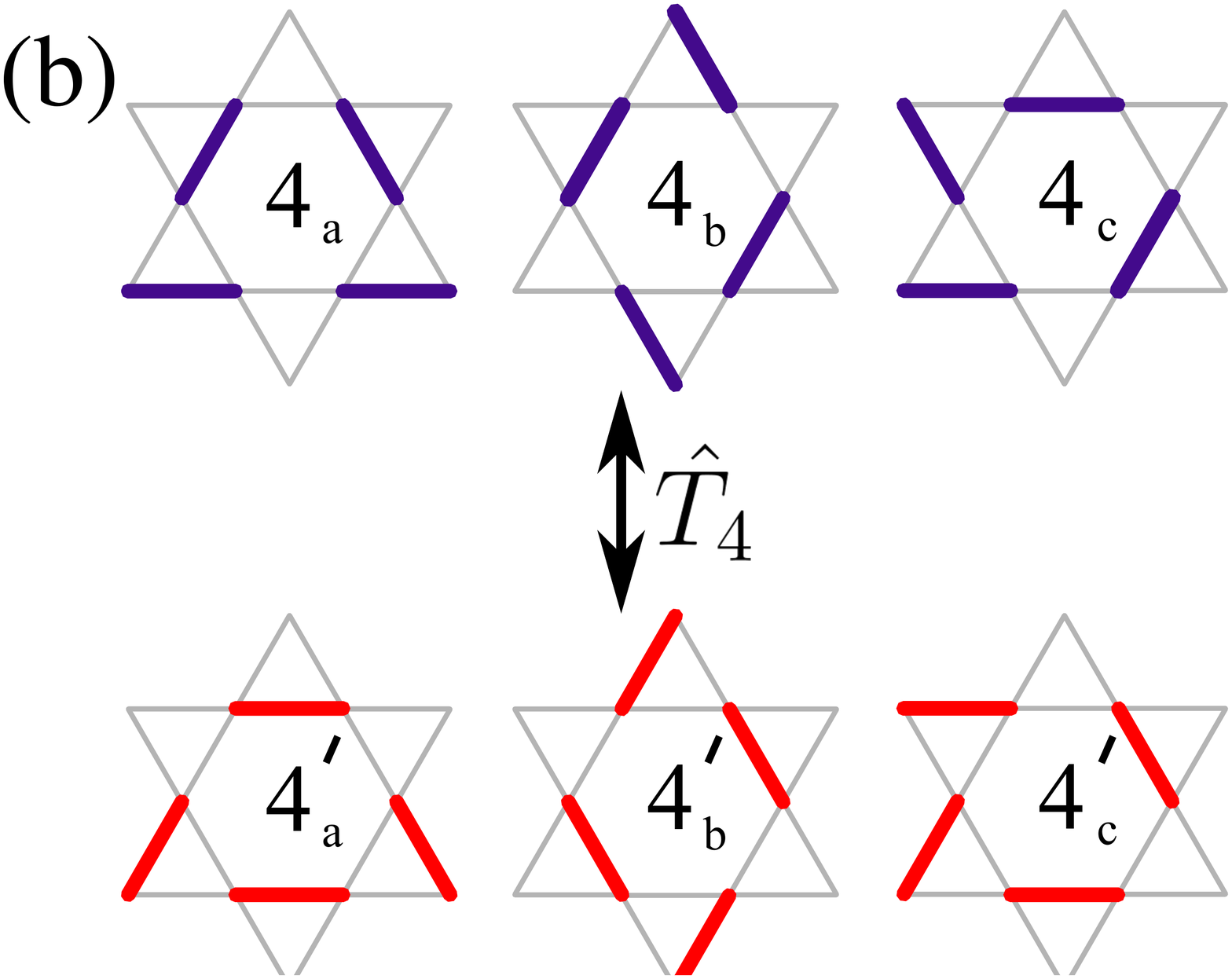}
\caption{The kagome lattice and the four-dimer plaquettes. In the upper panel (a), the thick purple bonds are dimers. The number at the centers of plaquettes indicate the number of dimers around the plaquette. The thin dashed line circles a cluster of seven plaquettes without a four-dimer plaquette. The lower panel (b) displays three types (a,b,c) of plaquettes with four dimers. Upon an application of $\hat{T}_4$ operator, the purple thick dimer configuration $4_{\mathrm{a,b,c}}$ become into the red thick dimer configurations $4^\prime_{\mathrm{a,b,c}}$ respectively. }\label{fig:kagome}
\end{figure}

\section{Results}
\subsection{Quantum Dimer Model on the kagome lattice} We proceed to construct our QDM. Taking a different approach from previous studies \cite{Zeng.1995,misguich.2003,Poilblanc.2010} which employ a QDM derived directly from the spin model, we follow an alternative strategy and study the  {minimal} QDM on the kagome lattice. Since the kinetic energy of collective motion decreases exponentially as the number of dimers involved increases \cite{Zeng.1995,misguich.2003,Poilblanc.2010}, the simplest dimer model on the kagome lattice includes only the motion and interaction of three dimers (Figure~\ref{fig:kagome}). However, the three-dimer motion does not change the number of ``perfect hexagons'', a hexagon of six triangles with only three dimers. Henceforth, we will refer to the hexagonal loops of six triangles as ``plaquettes''. The model is  {classical} since the kinetic term and the potential term commute. The lowest energy state of the classical model is any dimer configuration with a maximum (or minimum) number of perfect hexagons. The ground state is thus highly degenerate. The degeneracy is lifted by the lowest order quantum terms: kinetic energy and interaction of four dimers. Following these arguments, the minimal QDM on the kagome lattice is:
\begin{equation}\label{eq:hami}
H=-t \sum_{h}\hat{T}_4(h)+v \hat{N}_4+\alpha \left({ \sum_{h}t^\prime \hat{T}_3(h)+ v^\prime\hat{N}_3 }\right). 
\end{equation}
Here, $\hat{T}_n(h)$ moves $n$ dimers collectively around the plaquette $h$. $\hat{N}_n$ count the number of plaquettes with $n$ dimers.  In the limit of $\alpha\to\infty$, we recover the classical dimer model. In this work, we focus on the opposite quantum limit with $\alpha=0$:
\begin{equation}\label{eq:hami2}
H=-t \sum_{h}\hat{T}_4(h)+v \hat{N}_4\equiv t\left(-\sum_{h}\hat{T}_4(h)+\lambda \hat{N}_4\right)
\end{equation}
where $\lambda\equiv v/t$. 

We now prove, by  {reductio ad absurdum}, that  {any} dimer covering state on the kagome lattice is connected to some other states through the Hamiltonian \eqref{eq:hami2}. We note that this is only the  {necessary} condition for the ergodicity of our dimer model. Consider a kagome lattice with $M$ sites. The number of plaquettes with three, four, five and six dimers ($n_{l}$ with $l=3,4,5,6$) satisfy the following relations: 
\begin{equation}
\sum_{l=3}^{6} n_{l}=\frac{M}{3},\qquad
\sum_{l=3}^{6} l n_{l}= \frac{3M}{2}. 
\end{equation}
The average number of dimer per plaquette is $\bar{n}=4.5$, the exact median of $l=3\ldots 6$. If there is a dimer covering state with no four-dimer plaquettes, the state must have some three-dimer plaquettes. Due to geometrical constraints, only plaquettes with four or five dimers can be adjacent to a three-dimer plaquette. Assuming that there is no four-dimer plaquettes, the average dimer density of the cluster of seven plaquettes (Figure \ref{fig:kagome}) is $33/7\approx 4.7>\bar{n}$. In order to lower the average number of dimers to $\bar{n}$, the cluster must be adjacent to at least one perfect hexagon. This is geometrically impossible.  As a result, every dimer covering state must have a non-zero number of four-dimer plaquettes. 

\begin{figure}
\centering
\includegraphics[width=\columnwidth]{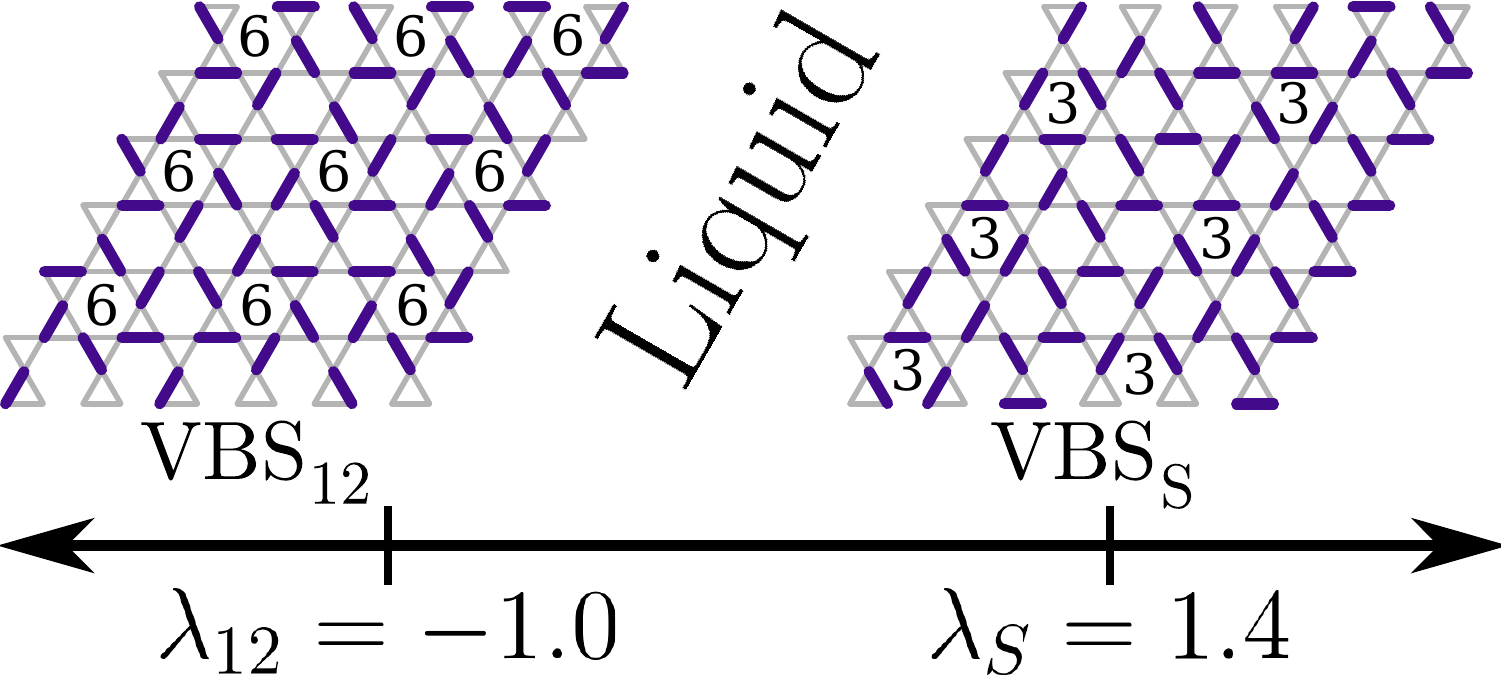}
\caption{The phase diagram of Hamiltonian \eqref{eq:hami2}. The dimer configurations of both $\text{VBS}_{12}$ (the 12-site VBS) and $\text{VBS}_{\text{S}}$ (the stripe VBS) are produced from the snapshots of QMC simulation for $L=6$. The purple thick bonds are the dimers. }\label{fig:diagram}
\end{figure}

The observation that $\bar{n}=4.5$ helps us anticipate the phase diagram of \eqref{eq:hami2} (Figure \ref{fig:diagram}). At $\lambda=1$, the Rokhsar-Kivelson point \cite{Rokhsar.1988}, the ground state of Hamiltonian \eqref{eq:hami2} is exactly known. It is a Z$_2$ liquid with short-range correlation \cite{misguich.2002}. At $\lambda$ increases, the four-dimer plaquettes are ``expelled'' from the ground state due to higher energy cost. To keep $\bar{n}=4.5$, the number of perfect hexagons increases. At large $\lambda$, one expects the ground state to be a valence-bond crystal with maximum number of perfect hexagons. There are two candidate states discussed in previous studies, the 36-site VBS \cite{singh.2008,Nikolic.2003} and the stripe VBS \cite{Nikolic.2003}, referred to as VBS$_{36}$ and VBS$_{\mathrm{S}}$ hereafter. Similarly, for $\lambda$ smaller than some critical $\lambda_{\mathrm{c_1}}$, the ground state is expected to be a valence-bond crystal with maximum number of four-dimer plaquettes, the 12-site valence-bond solid or VBS$_{12}$. We note that Yan  {et al} \cite{Yan.2011} demonstrated that the VBS$_{12}$ state is close in phase space to the Z$_2$ quantum spin liquid in their seminal DMRG work. 

\subsection{Quantum Monte Carlo simulation} We perform our numerical QMC simulations on finite kagome-lattice rhombi with $L\times L \times 3$ sites. Periodical boundary conditions are applied along two directions: $\hat{e}_1=\hat{x}$ and $\hat{e}_2=\hat{x}/2+\sqrt{3}\hat{y}/2$. The torus preserves all point group symmetry of the lattice \cite{Ralko.2005,Ralko.2006,Ralko2}. The four-fold degenerate ground states are labelled by two binary winding numbers $(\sigma_1,\sigma_2)$ which are the parities of the numbers of dimers crossed by a cut through the system along $\hat{e}_1$ and $\hat{e}_2$ directions, respectively. Assuming $L=2N$, either $(0,1)$ $(1,0)$ and $(0,0)$ or $(0,1)$ $(1,0)$ and $(1,1)$ states are related by six-fold rotational symmetries if $N$ is odd or even, respectively \cite{Ralko.2005,Ralko2}. To expose the topological property of the liquid state, we start the QMC simulation from initial configurations with inequivalent $(\sigma_1,\sigma_2)$. On a finite kagome rhombus, a particular VBS is compatible with some $(\sigma_1,\sigma_2)$ while the liquid state exists for all possible topological sectors. As a result, we expect the ground state energies to be degenerate among all topological sectors if the ground state is a Z$_2$ liquid. In contrast, the ground state energies of  the topological sectors will be different if the system orders into a VBS. 

\begin{figure}
\centering
\includegraphics[width=\columnwidth]{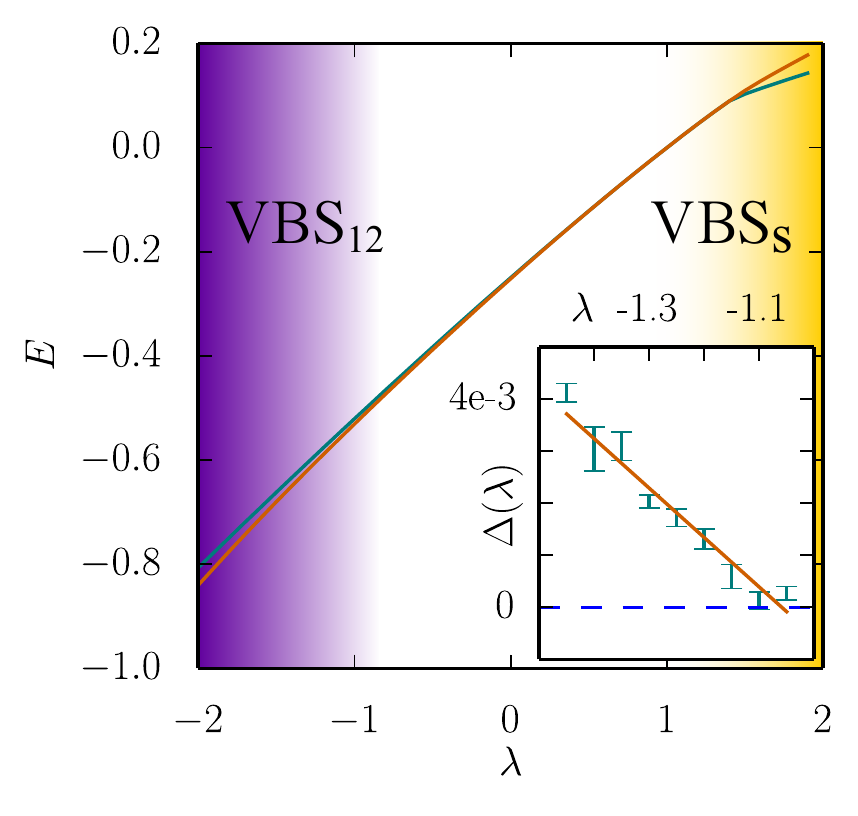}
\caption{The ground state energy for different topological sectors for $L=6$. The splitting of ground state energy as a function of $\lambda$ for topological sectors $(0,0)$ (the teal line)  and $(1,1)$ (the orange line).
The background gradient shows the onset of the two solid phases.
Inset: The energy splitting $\Delta(L)$ in the limit of $L\to \infty$ as a function of $\lambda$ (axis labels on top). }\label{fig:split}
\end{figure}

We first consider the torus with $L=6$. The rhombus is commensurate with the unit cells of the three VBS's. The ground state energy per plaquette is calculated as the function of $\lambda$ for both $(0,0)$ and $(1,1)$ sectors (Figure \ref{fig:split}). Approximately for $-1<\lambda<1.4$, the ground state energies of the two topological sectors are degenerate. This indicate a large stable region of Z$_2$ liquid phase of size $\delta \lambda\sim 2.4$ comparing to the $\delta \lambda \sim 0.3$ in the triangular lattice QDM \cite{Ralko.2005,Moessner.2001}. For $\lambda>1.4$, the system order into a VBS indicated by the abrupt splitting of the groundstate energies of the two topological sectors (Figure \ref{fig:split}). The dimer configuration from our simulation indicates that this is the VBS$_\mathrm{S}$ phase (Figure \ref{fig:diagram}). Similarly, for $\lambda<-1$, our model orders into VBS$_{12}$ evidenced in both the gradual splitting of the ground state energies, and real-space images of dimer configurations (Figure \ref{fig:diagram}) sampled from the simulation. 

First, we focus on the phase transition into VBS$_{\mathrm{S}}$. The abrupt nature of the energy splitting at $\lambda_{\mathrm{c_2}}\sim 1.4$ indicates that there is no VBS order for $\lambda<\lambda_{\mathrm{c_2}}$ in a finite system. Furthermore, We observe metastability of the liquid phase for $\lambda>\lambda_{\mathrm{c_2}}$. We tentatively conclude that the transition to VBS$_\mathrm{S}$ is first-order. 

Second, we consider the phase transition from the liquid phase to VBS$_{12}$. The energy splitting between topological sectors sets in slowly around $\lambda_{\mathrm{c_1}}\sim -1$ indicating that the solid order onsets in the finite $L=6$ system in a gradual fashion. This suggests that the transition could be continuous. We estimate $\lambda_{\mathrm{c_1}}$ in the $L\to \infty$ limit through scaling analysis of energy splitting $\delta E(\lambda,L)$ as a function of $L$. We calculate $\delta E(\lambda, L)$ under a fixed set of $\lambda$ for $L=6,8\ldots 16$. The data for different system sizes can be fitted to the following formula: $\delta E(\lambda, L)=\Delta(\lambda)+\frac{a(\lambda)}{L^2}$. $\Delta(\lambda)$  (Figure \ref{fig:split}) is the energy difference between topological sectors in the limit $L\to \infty$. While the scaling behavior of $\Delta(\lambda)$ is  {not} rigorously known around the phase transition point,  a simple linear fit (Figure \ref{fig:split}), which is equally justified, of $\Delta$ vs $\lambda$ suggests that $\lambda_{\mathrm{c_1}}\approx -1.06(1)$.  

\subsection{Critical properties and finite size scaling} To investigate the properties of the phase transition quantitatively, we measure the order parameter and the equal-time two-point correlator \cite{sachdev2001} of the order parameter in our QMC using the forward walking technique \cite{Buonaura.1998}. To define the order parameter, note that the VBS$_{12}$ state is a superposition of three dimer density-waves at the following wave vectors:
\begin{equation}
\mathbf{Q}_1=\frac{2\pi}{\sqrt{3}}\hat{y},\qquad \mathbf{Q}_{2,3}=\frac{2\pi}{\sqrt{3}}\left(\pm\frac{\sqrt{3}\hat{x}}{2}-\frac{\hat{y}}{2}\right). 
\end{equation}
The centers of all plaquettes,  labeled as ${\bm x}$, form a triangular lattice. We define $n({\bm x})$ to be the total number of dimers around plaquette ${\bm x}$, while $m_i$ with $i=1,2,3$ are defined as modulations of dimer density at $\mathbf{Q}_i$:
\begin{equation}\label{eq:mi}
m_{i}=\frac{1}{L^2}\sum_{\bm x}(n({\bm x})-\bar{n})\exp(i \mathbf{
Q}_i\cdot{\bm x}). 
\end{equation}
We note that $\mathbf{Q}_{i}=-\mathbf{Q}_{i}$ ($i=1,2,3$). As a result, $m_i$'s are real numbers. The components $m_i$ form a three dimensional vector, $\mathbf{m}=(m_1,m_2,m_3)$, which is the proper order parameter defining the VBS$_{12}$ order in the thermodynamic limit. The space group symmetries guarantee that, in the ground state wave function of a finite system, the weights for dimer configurations with $\mathbf{m}$ of opposite  {directions} but the same magnitudes are equal.  As a result, for any finite system the expectation value of $\mathbf{m}$ with respect to the ground state wave function is strictly zero. In order to characterize the VBS order in finite systems, we thus define the order parameter $m\equiv |\mathbf{m}|$ and the corresponding equal-time two-point correlator $\chi$ \cite{Sandvik.2010}:
\begin{subequations}\label{eq:oands}
\begin{eqnarray}
m&\equiv& \sqrt{m_1^2+m_2^2+m_3^2}, \label{eq:opara} \\
\chi&\equiv &(\langle m^2\rangle-\langle m\rangle^2)L^2. 
\end{eqnarray}
\end{subequations}
If the system has the saturated VBS$_{12}$ order, $\langle m\rangle =\sqrt{3}/2\approx 0.866$. This is the maximum value of the order parameter $m$. 

\begin{figure}
\centering
\includegraphics[width=0.9\columnwidth]{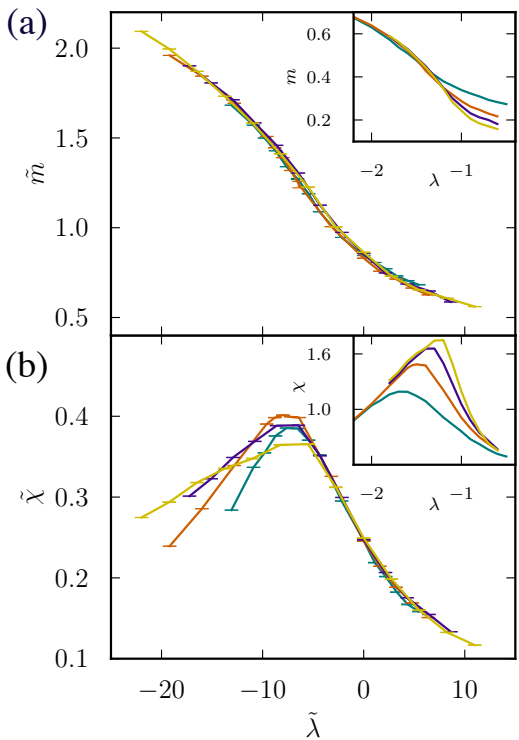}
\caption{The finite-size scaling of the order parameter and the equal-time two-point correlator.
The order parameter (a) and the equal-time two-point correlation (b) for different system sizes collapse onto universal curves under proper scaling. $L=6, 8,10,12$ are represented by teal, orange, purple and yellow lines. Extensive order parameter (upper inset) and equal-time two-point correlator (lower inset) show how the jump in the order parameter and the peak equal-time correlator change with system size.
}\label{fig:collapse}
\end{figure}

We measure both $m$ and $\chi$ as functions of $\lambda$ for $L=6,\ldots 12$. 
For all system sizes, the order parameter shows a continuous increase from zero  toward the saturation value at $\lambda\ll \lambda_{\mathrm{c_1}}$.
As $\lambda$ becomes smaller, the values of order parameters for different system sizes converge onto the same value, which is typical behavior for a continuous phase transition.
To measure the critical exponents, we plot the data for different system sizes using the following scaling form \cite{Sandvik.2010}:
\begin{equation}\label{eq:opscaling}
m=L^{-\frac{\beta}{\nu}}F\left(\frac{(\lambda-\lambda_{\mathrm{c_1}})L^{\frac{1}{\nu}}}{\lambda_{\mathrm{c_1}}}\right)
\end{equation}
where $F(x)$ is the universal scaling function.
For $\lambda_{\mathrm{c_1}}=-1.00(6)$, $\beta/\nu=0.51(4)$ and $\nu=0.75(6)$, all of our data collapses on a universal curve (Figure~\ref{fig:collapse}). 
The success of this scaling is very strong evidence that the phase transition is indeed continuous. 

For $\chi$ ($L=6\ldots 12$), 
we observe that $\chi(\lambda)$ show a peak for all system sizes.
As $L$ increase, the peak shifts right while its height increase.
Away from the peak position, the $\chi$ data converges to the same value for all system sizes.
For a conventional continuous phase transition in two dimension, the peak height should increase approximately as $L^2$ due to the small value of anomalous dimension $\eta$ in conventional universality classes \cite{Campo}.
In contrast, the peak height at $L=12$ is less than twice of the peak height for $L=6$, signalling a large $\eta$.
We assume the following scaling form for $\chi$ \cite{sachdev2001,Sandvik.2010}:
\begin{equation}\label{eq:chiscaling}
\chi=L^{2-\eta}G\left(\frac{(\lambda-\lambda_{\mathrm{c_1}})L^{\frac{1}{\nu}}}{\lambda_{\mathrm{c_1}}}\right). 
\end{equation}
Here $G(x)$ is the universal scaling function.
Optimizing close to $\lambda_{\mathrm{c_1}}$, all our data collapse on a universal curve with $\nu=0.75(6)$ and $\eta=1.37(8)$ (Figure~\ref{fig:collapse}).
Obtaining $\eta$ to higher accuracy is impossible due to the decreasing quality of data collapse away from the critical region, and the 
difficulty in fitting the peak of the largest $L$;
regardless, the existing data provides compelling evidence for a large anomalous dimension $\eta$.
This is the definitive signature of an unconventional quantum phase transition \cite{Xu_1} induced by the condensation of nonlocal excitations -- Z$_2$ vortices or visons. 
In other words, the physical interpretation is that singlet density fluctuations are not the coherent excitations in a Z$_2$ liquid;
they fractionalize into pairs of visons.
The equal-time two-point correlator that we measure via QMC correspond to the tensorial correlator of four visons. 

\section{Discussion}
The critical exponents we obtain are consistent with the exponents of the O$(4^\ast)$ universality class \cite{Xu_1}:  $\beta/\nu=0.5129(11)$ \cite{Kanaya.1995}, $\nu=0.7525(10)$ \cite{Ballesteros} and $\eta=1.373(3)$ \cite{Isakov.2005}. 
An effective field theory study \cite{Huh.2011} indicates that the critical theory of the transition to VBS$_{12}$ involves four components of real fields. However, the most general quartic term does not respect O$(4)$ symmetry \cite{Huh.2011}.  While a two-loop perturbative renormalization group (RG) study \cite{Toledano.1985} does not reveal a stable fixed point, it is possible that a higher-order RG study could yield a stable fixed point, indicating a generic second-order phase transition. The potential new fixed point would generally possess critical exponents different from the O$(4^\ast)$ universality class. However, the numerical difference could be small and therefore not appreciated in our current simulation.

While in this paper we studied a QDM with general parameters, it is likely that our liquid phase is adiabatically connected to the Z$_2$ quantum spin liquid phase identified in previous DMRG studies \cite{Yan.2011,Depenbrock.2012,Jiang2012} on the kagome lattice. The kinetic and interaction energies for three and four valence bonds are the largest in the QDM \cite{Zeng.1995,misguich.2003,Poilblanc.2010} derived from the spin model. We have demonstrated that the three valence-bond terms generically favor frozen VBS states. The four valence bond terms are thus the quantum fluctuations that melt the VBS states and induce the liquid phase.  Previous studies \cite{Yan.2011} also indicate that the Z$_2$ liquid phase is close to VBS$_{12}$. Given that the phase transition from the liquid to the VBS is likely to be continuous, the singlet excitations (pairs of visons) can have a very small energy cost. This is consistent with the large number of singlet excitations within the spin-gap observed in exact diagonalization studies \cite{Waldtmann.1998}. It is also consistent with the difficulty in pinpointing the minimal energy cost to create singlet excitations in DMRG calculations \cite{Yan.2011}. 

We conclude by commenting on the relevance of our results on real physical systems. 
Our unconventional transition is reliant upon visons condensing out of a Z$_2$ ``parent'' phase, of which there are very few candidates in real
material systems \cite{LeonQSL}.   
However, it appears that elementary Z$_2$ states will soon be fabricated in superconducting quantum circuits \cite{DiVincenzo,Fowler1,transmon,transmon2},
raising the possibility of realizing interesting many-body physics in artificial quantum systems.
Indeed, there have been proposals in the past \cite{Ioffe.2002} to use Josephson junction arrays to realize a Z$_2$ liquid phase in the triangular lattice QDM \cite{Moessner.2001}. In that model, the maximum gap to low energy excitations is only about $0.1t$ at the RK point \cite{Ioffe.2002}. As a result, the liquid phase exists only for a small parameter space \cite{Ralko.2005,Moessner.2001} and any practical realization would have to be fine-tuned.  However, for our present model, this gap is likely on the order of $t$ evidenced by the larger range of parameter space where the liquid phase prevails. This certainly improves the prospect of realizing our QDM in an artificial system such as a Josephson Junction array. It has also been shown that the signature of the unconventional quantum phase transition, i.e~the large $\eta$, can be accessed experimentally, for example, through a spin-lattice relaxation time $T_1$ scaling with temperature $T$ in nuclear magnetic resonance experiments \cite{Yang.2009}. 
The fact that novel concepts such as fractionalization, deconfinement, and unconventional quantum criticality -- usually discussed in the context of 
condensed matter theory -- may first be manifested in artificial quantum systems should motivate the engineering and fabarication effort of these systems in the near future.


\section{Acknowledgements}
We thank M. Mariantoni and Michel J. P. Gingras for useful discussions, and acknowledge crucial correspondence with S. Sorella about the GFMC method and the forward walking technique in particular. This research was supported by NSERC of Canada, the Perimeter Institute for Theoretical Physics, and the John Templeton Foundation.  Research at Perimeter Institute is supported through Industry Canada and by the Province of Ontario through the Ministry of Research \& Innovation. Numerical simulations are carried out on the shared hierarchical academic research computing network commonly known as SHARCNET. 

\section{Author contributions}
ZH conceived the project, wrote the QMC code. SI and ZH performed the simulation and analyzed data with input from RM. All authors contributed to interpretation of the results and writing of the manuscript. 

The authors declare no competing financial interests.  
\bibliography{biblio}

\end{document}